\documentclass[pre,onecolumn,showpacs,preprintnumbers,pra,showkeys,amssymb]{revtex4}
\usepackage{amsfonts}
\usepackage{amsmath}
\usepackage{graphicx}
\usepackage{dcolumn}
\usepackage{epsfig}
\usepackage{color}
\usepackage{hyperref}
\usepackage{hyphenat}
\usepackage{bm}
\usepackage{float}
\usepackage{hyperref}
\usepackage{hyphenat}
\usepackage{color}
\usepackage{makeidx}
\usepackage{index}
\newcommand{\be}{\begin{equation}}
\newcommand{\beast}{\begin{equation*}}
\newcommand{\ee}{\end{equation}}
\newcommand{\eeast}{\end{equation*}}
\newcommand{\br}{\begin{eqnarray}}
\newcommand{\brast}{\begin{eqnarray*}}
\newcommand{\er}{\end{eqnarray}}
\newcommand{\erast}{\end{eqnarray*}}
\newcommand{\bse}{\begin{subequations}}
\newcommand{\ese}{\end{subequations}}

\newcommand{\bd}{\begin{displaymath}}
\newcommand{\ed}{\end{displaymath}}

\newcommand{\bfig}{\begin{figure}}
\newcommand{\efig}{\end{figure}}

\newcommand{\aspe}{\textquotedblleft}
\newcommand{\aspd}{\textquotedblright}
\makeindex
\begin{document}
\title{Urban vehicular traffic: fitting the data using a hybrid stochastic model. Part II}
\author{Ariel Amadio$^{1}$, Facundo Nicuesa$^{1}$, D. Otero$^{1}$, D. Galetti$^{2}$, and S. S. Mizrahi$^{3}$}
\email{ariel.amadio.aaa@gmail.com, n_facundo@hotmail.com, dinootero@fibertel.com.ar, diogaletti@hotmail.com, salomon@df.ufscar.br}$^{1}$
\affiliation{$^{1}$Facultad Regional General Pacheco, Universidad Tecnol\'{o}gica Nacional (UTN), 
Buenos Aires, Argentina \\
$^{2}$Instituto de F\'{\i}sica Te\'{o}rica, UNESP, S\~{a}o Paulo, SP, Brasil  \\
$^{3}$Departamento de F\'{\i}sica, CCET, Universidade Federal de S\~ao
Carlos, 13565-905, S\~ao Carlos, SP, Brazil}
\date{\today}

\begin{abstract}
In this second part of our research we used the models presented in 
\emph{Modeling a vehicular traffic network. Part I} \cite{ogm1} to 
perform an analysis of the urban traffic as recorded by cameras 
distributed in a chosen sector of Tigre, a city in the province of Buenos 
Aires, Argentina. We found that the circulation of vehicles -- the traffic 
dynamics --, along a whole day, can be described by a hybrid model that is 
an adapted blend of model 2, for an open linear system, with model 3, which 
is nonlinear, developed in Part I. The objectives of this work were, firstly, 
to verify whether the vehicular flux can be modeled as an $n$-step stochastic 
process for its evolution, $n$ for the time. Secondly, to find out if 
the model, with its parameters fixed to describe the traffic of a single day, 
may adequately describe the traffic in other days. Thirdly, to propose changes 
in the already established set of the urban traffic rules in order to optimize 
the vehicular flow and to diminish the average time that a vehicle stays idle 
at the semaphores. We estimate that the goals were achieved satisfactorily 
within the margins of the experimental errors of the gathered data.  
\end{abstract}

\maketitle




%
%
\section{Introduction}
\begin{figure}[htbp]
\centering
\includegraphics[height=5.0 in, width=3.6 in]{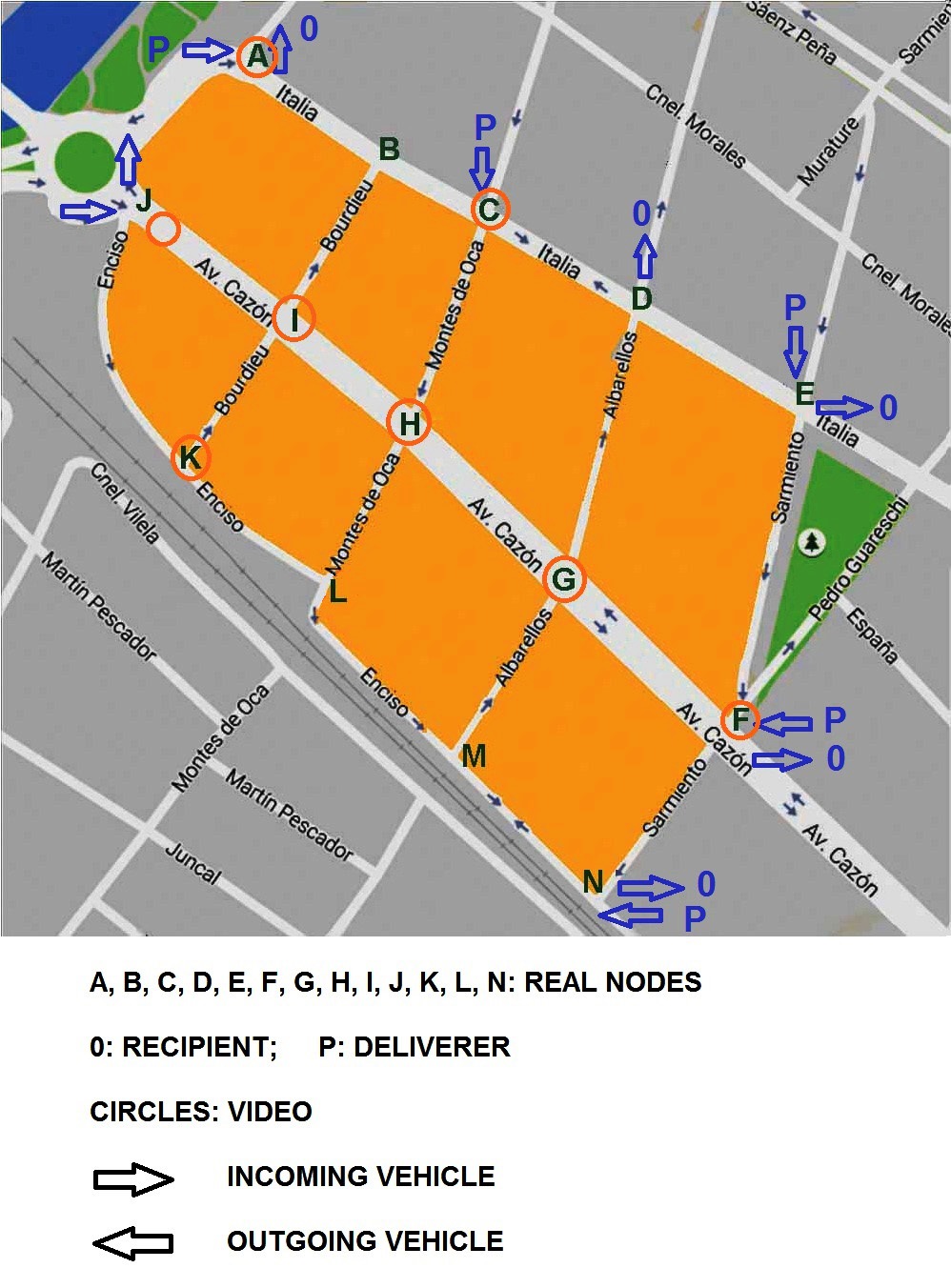}
\caption{{\small{Chosen sector in downtown of Tigre. It consists
of fourteen intersections, with cameras installed in eight of them 
(identified by the orange circles) for recording the traffic flow.}}}
\label{sector}
\end{figure}

In a preceding paper, to be referred as Part I (\emph{Modeling a vehicular 
traffic network. Part I} \cite{ogm1}), we developed three models for the 
description of networks representing the vehicular traffic between several sites 
that could be cities, parking lots, car rental agencies, etc...\ . A network 
is composed by arteries, intersections and blocks; pictorially, it is represented by 
a digraph made of vertices and directed edges. Mathematically the digraph is 
pictured as a matrix, whose diagonal entries represent the number of vehicles 
at the sites (vertices) and the off-diagonal ones stand for the number of vehicles 
in transit, from one site to another, along the arteries (edges). By normalizing 
each row, one gets a stochastic matrix, whose entries are numbers in the interval 
$\left[0,1\right]$. The evolution of a vehicular distribution at time $n$ is 
obtained step by step: one multiplies the initial vector, whose components stand 
for the distribution of vehicles within the network at $n=0$, by the stochastic 
matrix raised to the power $n$. This procedure is admitted to be adequate to 
describe the behavior of the traffic dynamics, as long as that matrix has been 
constructed confidently. In Part I we have presented a theoretical description, 
with numerical examples, of a finite network with linear and nonlinear models 
for closed and open systems. In the present Part II we used the methods and 
models of Part I to analyze the collected data for the traffic of vehicles in 
a chosen sector of the city of Tigre, Province of Buenos Aires, Argentina, 
see Fig. \ref{sector}. The detailed formalism is presented in \cite{ogm1} and 
in the therein references.

We constructed two stochastic matrices, $\mathbb{M}^{(1)}$ and $\mathbb{M}^{(2)}$, 
one for each period of a particular day (from 06:00 AM to 08:00 PM and from 08:00 
PM to 06:00 AM of the following day), that when evolved, according to the proposed 
model, permitted to verify that the observed vehicular distribution is reproduced 
satisfactorily within the experimental error of the data. Even for other days 
the vehicular distributions were adequately described using the very same matrices 
$\mathbb{M}^{(1)}$ and $\mathbb{M}^{(2)}$. 

Since the vertices of a digraph represent street intersections in the city map, 
we will also refer to them as nodes. We shaped the dynamics of the influx and 
outflux of vehicles by choosing conveniently the entries of $\mathbb{M}^{(1)}$ 
and $\mathbb{M}^{(2)}$. We found that a hybrid model that is an adapted merging 
of the linear open system model with the nonlinear one, introduced in the Part I, 
produces numerical results that are in fair agreement with the data from the 
traffic flow. The data we used came from samples counts although the traffic was 
continuously monitored during 24 hours, . 

The traffic in a city, and even in one of its sectors, is not uniform 
during a whole 24-hour day, instead it varies continuously. In our chosen 
sector we verified that for an usual working day the influx begins 
to increase steadfastly after 6:00 AM. Then, from 10:00 AM to 8:00 PM it 
stabilizes with some oscillations, that we call a quasi-stationary r\'{e}gime, 
which is due to the commercial and social activities, namely, the opening 
and closing of businesses, shops, offices, hospitals, police station and schools 
within the perimeter of the network. Afterward it decreases rapidly, 
the vehicular outflux starts at 8:00 PM (when the activities reduce at 
the end of the day) and continues until 4:00 AM of the following day. From 4:00 AM 
to 6:00 AM there is practically no circulation of vehicles. We are interested 
in modeling this dynamics.  

\begin{figure}[htbp]
\centering
\includegraphics[height=2.1in, width=3.6 in]{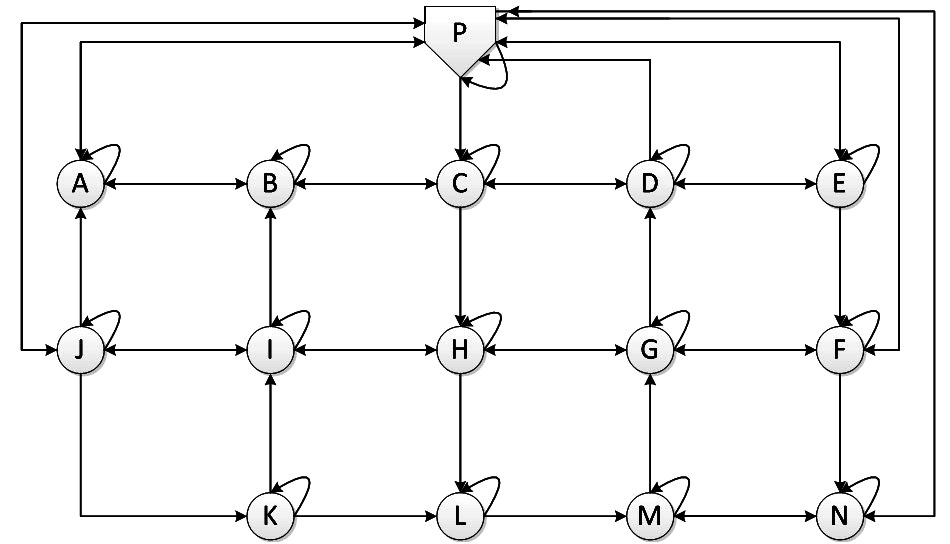}
\caption{{\small{Digraph picturing the network. The node $P$ does not 
represent a real intersection, it is an extra auxiliary node (a reservoir) 
accounting for the influx and outflux of vehicles within the network. All 
other nodes represent the arteries intersections. Each of them has a loop 
to which one associates the number of vehicles standing idle at 
red lights or parked along the sidewalks.}}}
\label{netdigraph}
\end{figure}

The selected sector of the city, see Fig. \ref{sector}, has several blocks 
separated by streets and avenues, where the arrows indicate the directions 
of the vehicular flux. Altough the cameras were positioned at only eight 
intersections, labeled $A,C,F,G,H,I,J,K$ and marked by the orange circles, 
we have included six additional intersections, thus resulting in a network of 
fourteen intersections. The roads and avenues that link them are referred 
by their names. This sector has a reduced number of entries and exits, and 
still conveys a heavy traffic during the day without the occurrence of 
significant traffic jam. The heavier traffic occurs in Caz\'{o}n avenue, 
because it is the artery where most businesses and public offices are located. 
Then comes Italia avenue, whereas the other arteries have comparatively 
less intense traffic. This network is an open system, because at the nodes 
$A,C,E,F,N$ and $J$ the vehicles arrive from outside the circuit (an influx 
of vehicles), while at nodes $A,D,E,F,N$ and $J$ there is a drain of vehicles 
(an outflux). According to the observed characteristics of the traffic, we are 
modeling the system assuming that, basically, the vehicles circulating within 
the network arrive at and depart through the nodes, with the exception of a small 
number that are parked along the sidewalks nearby the intersections plus those 
standing idle awaiting for the semaphores to change light. An extra auxiliary 
node $P$ was added to the circuit, which assumes the r\^{o}le of a reservoir, 
it represents a locus of influx and outflux of vehicles for the network, see 
Figs. \ref{sector} (map) and \ref{netdigraph} (digraph). Recently, another 
model \cite{manley}, quite different from ours, was proposed with the same aim,  
to describe the vehicular traffic flow.
%
\section{Hybrid model for vehicular traffic}
%
The evolution of the average distribution of vehicles in our network is represented 
by the vector $\mathbf{v}^{\dagger }\left( n\right) $, with $n$ a positive integer, 
standing for discretized time, as
\begin{equation}
\mathbf{v}^{\dagger }\left( n\right)  =\mathbf{v}_{0}^{\dagger }%
\mathbb{M}_{eff}\left( n\right) ,  
\label{s1}
\end{equation}
where each component of $\mathbf{v}^{\dagger }\left( n\right)$ is the 
vehicular distribution at one node plus those in transit to other nodes, 
at time $n$, and $\mathbf{v}_{0}^{\dagger }$ stands for the initial distribution. 
The effective matrix is the product of $n+1$ matrices,
\begin{equation}
\mathbb{M}_{eff}\left( n\right) = \mathbb{E}\left( n\right) \mathbb{M}^{n}\ .  
\label{s1a}
\end{equation}
The matrix $\mathbb{M}$ is stochastic and its entries stand 
for the initial conditions for the evolution; the matrix $\mathbb{E}\left(n\right) $ 
contains the effects of the environment, because the vehicular traffic is an 
open system. $\mathbb{E}\left( n\right)$ is not stochastic and we call it 
the \emph{environment influence matrix}. The effective matrix (\ref{s1a}) is 
the generator of the evolution of $\mathbf{v}_{0}^{\dagger }$. In short, the 
evolution of an open nonlinear network is based on a hybrid of two models 
and the vehicular distribution $\mathbf{v}^{\dagger }\left( n\right) $ is also 
nonlinear as the sum of its components changes with $n$. As $\mathbb{M}$, the 
matrix $\mathbb{E}\left( n\right) $ is also a square matrix of dimension $N$, 
that we defined as
\begin{equation}
\mathbb{E}\left( n\right) =\left( 
\begin{array}{cccccc}
1 & 0 & \cdots & 0 & 0 & 0 \\ 
& 1 & \cdots & 0 & 0 & 0 \\ 
\cdots & \cdots & \ddots & 0 & \cdots & \cdots \\ 
0 & 0 & \cdots & 1 & 0 & 0 \\ 
0 & 0 & \cdots & 0 & 1 & 0 \\ 
0 & 0 & \cdots & 0 & 0 & 1+\frac{f\left( n\right) }{\mathcal{N}}%
\end{array}
\right)  \, ,
\label{s2}
\end{equation}
where $f\left( n\right) $ is a specific function and $\mathcal{N}$ is the
total number of vehicles within the virtual reservoir $P$. The parameters are 
adjusted by minimizing the $\chi ^2$ deviations from the collected data. 
We outline the network as the array in Table \ref{tab1}, where the numbers 
for non-zero entries are set either by the counts of vehicles, or by 
estimation for the intersections without cameras. The zeros stand for the 
wrong ways for the vehicles.
\begin{table}[htb]
\centering
\begin{tabular}{|c||c|c|c|c|c|c|c|c|c|c|c|c|c|c|c|}
\hline
& $A$ & $B$ & $C$ & $D$ & $E$ & $F$ & $G$ & $H$ & $I$ & $J$ & $K$ & $L$ & $M$
& $N$ & $P$ \\ \hline\hline
$A$ & $p_{AA}$ & $p_{AB}$ & $0$ & $0$ & $0$ & $0$ & $0$ & $0$ & $0$ & $%
p_{AJ} $ & $0$ & $0$ & $0$ & $0$ & $p_{AP}$ \\ \hline
$B$ & $p_{BA}$ & $p_{BB}$ & $p_{BC}$ & $0$ & $0$ & $0$ & $p_{BG}$ & $0$ & $0$
& $0$ & $0$ & $0$ & $0$ & $0$ & $0$ \\ \hline
$C$ & $0$ & $p_{CB}$ & $p_{CC}$ & $p_{CD}$ & $0$ & $0$ & $0$ & $p_{CH}$ & $0$
& $0$ & $0$ & $0$ & $0$ & $0$ & $0$ \\ \hline
$D$ & $0$ & $0$ & $p_{DC}$ & $p_{DD}$ & $p_{DE}$ & $0$ & $p_{DG}$ & $0$ & $0$
& $0$ & $0$ & $0$ & $0$ & $0$ & $p_{DP}$ \\ \hline
$E$ & $0$ & $0$ & $0$ & $p_{ED}$ & $P_{EE}$ & $p_{EF}$ & $0$ & $0$ & $0$ & $%
0 $ & $0$ & $0$ & $0$ & $0$ & $p_{EP}$ \\ \hline
$F$ & $0$ & $0$ & $0$ & $0$ & $0$ & $p_{FF}$ & $p_{FG}$ & $0$ & $0$ & $0$ & $%
0$ & $0$ & $0$ & $p_{FN}$ & $p_{FP}$ \\ \hline
$G$ & $0$ & $0$ & $0$ & $p_{GD}$ & $0$ & $p_{GF}$ & $p_{GG}$ & $p_{GH}$ & $0$
& $0$ & $0$ & $0$ & $0$ & $0$ & $0$ \\ \hline
$H$ & $0$ & $0$ & $0$ & $0$ & $0$ & $0$ & $p_{HG}$ & $p_{HH}$ & $p_{HI}$ & $%
0 $ & $0$ & $p_{HL}$ & $0$ & $0$ & $0$ \\ \hline
$I$ & $0$ & $p_{IA}$ & $0$ & $0$ & $0$ & $0$ & $0$ & $p_{IH}$ & $p_{II}$ & $%
p_{IJ}$ & $0$ & $0$ & $0$ & $0$ & $0$ \\ \hline
$J$ & $p_{JA}$ & $0$ & $0$ & $0$ & $0$ & $0$ & $0$ & $0$ & $p_{JI}$ & $%
p_{JJ} $ & $p_{JK}$ & $0$ & $0$ & $0$ & $p_{JP}$ \\ \hline
$K$ & $0$ & $0$ & $0$ & $0$ & $0$ & $0$ & $0$ & $0$ & $p_{KI}$ & $0$ & $%
p_{KK}$ & $p_{KL}$ & $0$ & $0$ & $0$ \\ \hline
$L$ & $0$ & $0$ & $0$ & $0$ & $0$ & $0$ & $0$ & $0$ & $0$ & $0$ & $0$ & $%
p_{LL}$ & $p_{LM}$ & $0$ & $0$ \\ \hline
$M$ & $0$ & $0$ & $0$ & $0$ & $0$ & $0$ & $p_{MG}$ & $0$ & $0$ & $0$ & $0$ & 
$0$ & $p_{MM}$ & $p_{MN}$ & $0$ \\ \hline
$N$ & $0$ & $0$ & $0$ & $0$ & $0$ & $0$ & $0$ & $0$ & $0$ & $0$ & $0$ & $0$
& $p_{NM}$ & $p_{NN}$ & $p_{NP}$ \\ \hline
$P$ & $p_{PA}$ & $0$ & $p_{PC}$ & $0$ & $p_{PE}$ & $p_{PF}$ & $0$ & $0$ & $0$
& $p_{PJ}$ & $0$ & $0$ & $0$ & $p_{PN}$ & $p_{PP}$ \\ \hline
\end{tabular} 
\caption{\small{The entries $p_{ij}$ are for the number of vehicles at vertices 
and edges associated with the digraph \ref{netdigraph} that represents the chosen 
city sector shown in Fig. \ref{sector}.}}
\label{tab1}
\end{table}
%
\section{Methodology of collecting and analyzing the data}
%
Our \emph{modus operandi} consisted in counting cars, vans, trucks, buses, 
and motorcycles. We adopted a sampling procedure: at each intersection having 
an installed camera we counted the number of vehicles during six consecutive 
minutes at every two hours. For example, beginning at 10:00 AM we counted until 
10:06 AM, which we adopted as a $5\%$ sample. 

At 6:00 AM and early hours of the morning the number of vehicles in circulation is
sparse, so we established the time-zero vector as $\mathbf{v}%
_{0}^{\dagger }=\left( \begin{array}{ccccc} 0 & 0 & \cdots  & 0 & \mathcal{N}\end{array}%
\right) $, having fifteen components with the first fourteen being null and
the fifteenth standing for the number of vehicles in node $P$. 
%

The distances that separate the intersections (block lengths) vary 
between $120$ and $140\ m$, and we verified that a vehicle takes 
between $20\ s$ and $25\ s$ to go from one intersection to the next. 
Therefore, we adopted $22.5\ s$ as the unit of time, i.e. for the 
transition from one node to the next for the $n$-step stochastic 
process. The evolution of the vehicular distribution within the 
network is directly associated with this time. For example, one hour 
corresponds to $160$ time units, thus the matrices 
(\ref{s8}) and (\ref{s10}) are raised to powers as 
$\left( \mathbb{M}^{(i)}\right)^{n}$, $n=1,..,160$, for their evolution 
along one hour.

\begin{figure}[htbp]
\centering
\includegraphics[height=3.2in, width=3.4in]{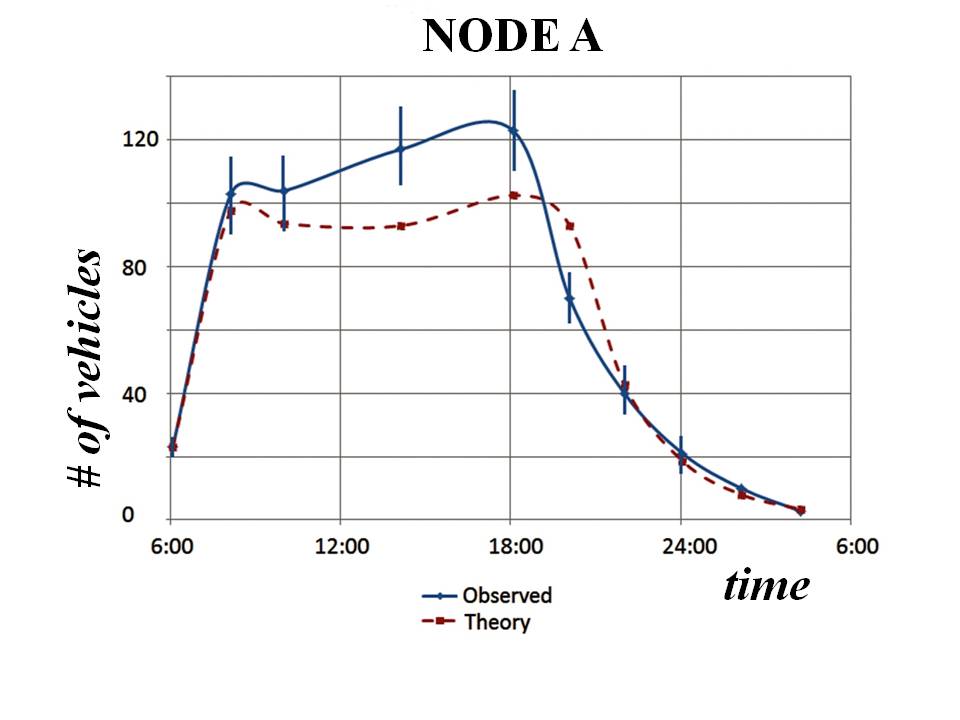}
\caption{\small{Variation of the flux (number of vehicles for every six 
minutes) going through the node $A$ along $22$ hours, 
beginning nearby 8:00 AM until 4:00 AM of the next day. Below the abscissa, 
the time in conventional notation. The error bars represent the dispersion of the 
empirical data collected in May 18, 2016. The hybrid model produces the dashed 
curve in red.}}
\label{p01}
\end{figure}

\begin{figure}[htbp]
\centering
\includegraphics[height=3.2in, width=3.4in]{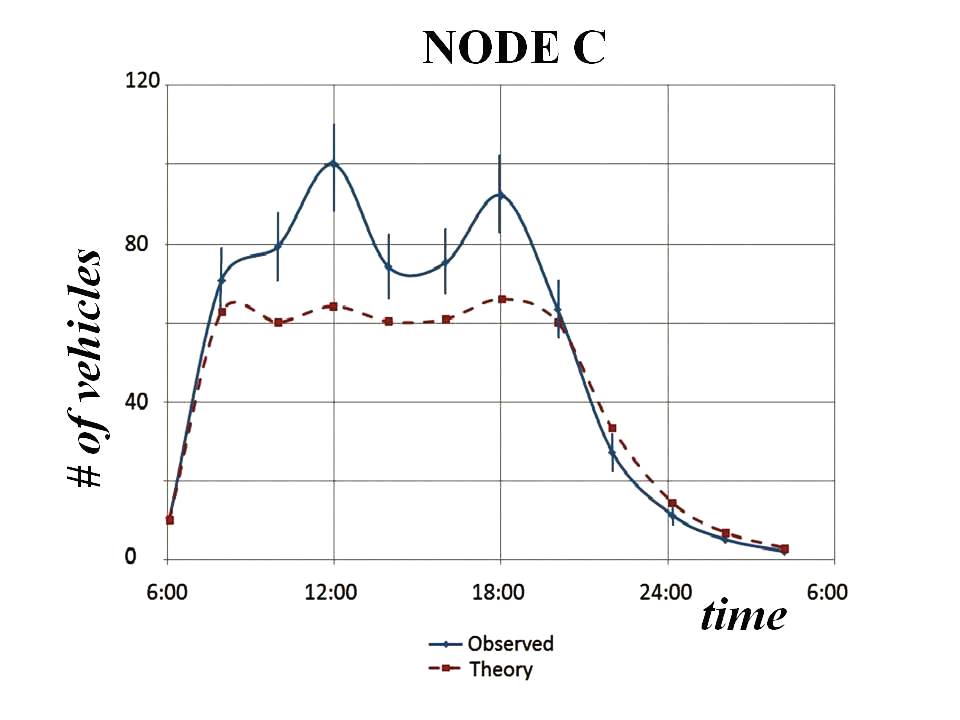}
\caption{\small{Variation of the flux (number of vehicles for every six 
minutes) going through the node $C$ along $22$ hours, 
beginning nearby 8:00 AM until 4:00 AM of the next day. Below the abscissa, 
the time in conventional notation. The error bars represent the dispersion of the 
empirical data collected in May 18, 2016. The hybrid model produces the dashed 
curve in red.}}
\label{p02}
\end{figure}

\begin{figure}[htbp]
\centering
\includegraphics[height=3.2in, width=3.4in]{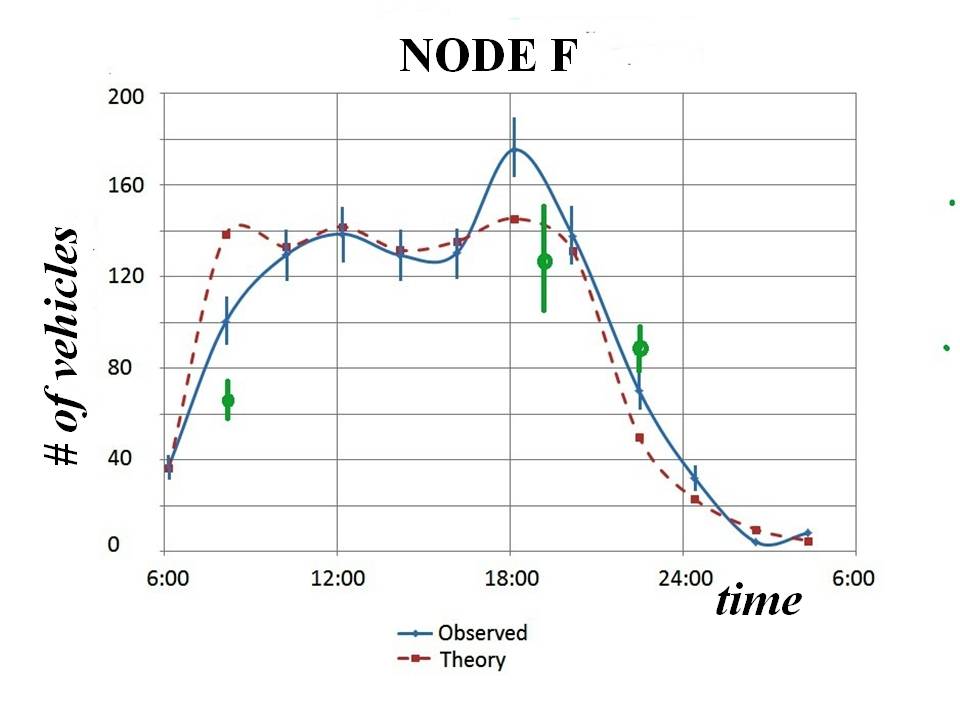}
\caption{\small{Variation of the flux (number of vehicles for every six 
minutes) going through the node $F$ along $22$ hours, 
beginning nearby 8:00 AM until 4:00 AM of the next day. Below the abscissa, 
the time in conventional notation. The error bars, in blue, represent the dispersion 
of the empirical data collected in May 18, 2016. The error bars, in green, represent 
the data collected in April 30, 2016. The hybrid model produces the dashed 
curve in red.}}
\label{p03}
\end{figure}

\begin{figure}[htbp]
\centering
\includegraphics[height=3.2in, width=3.4in]{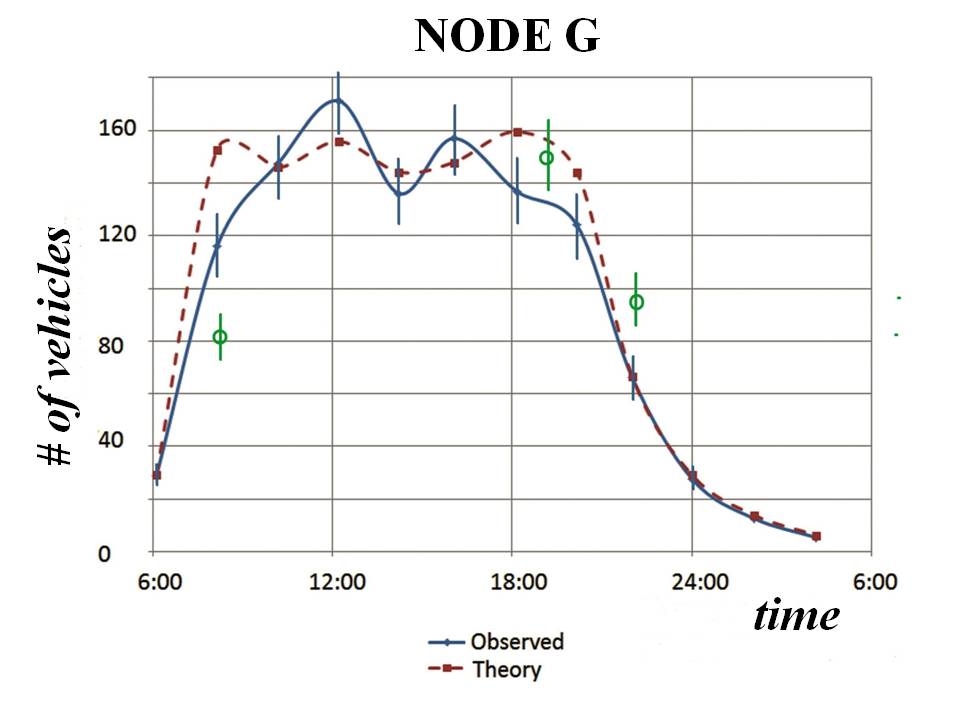}
\caption{\small{Variation of the flux (number of vehicles for every six 
minutes) going through the node $G$ along $22$ hours, 
beginning nearby 8:00 AM until 4:00 AM of the next day. Below the abscissa, 
the time in conventional notation. The error bars, in blue, represent the dispersion 
of the empirical data collected in May 18, 2016. The error bars, in green, represent 
the data collected in April 30, 2016. The hybrid model produces the dashed 
curve in red.}}
\label{p04}
\end{figure}

\begin{figure}[htbp]
\centering
\includegraphics[height=3.2in, width=3.4in]{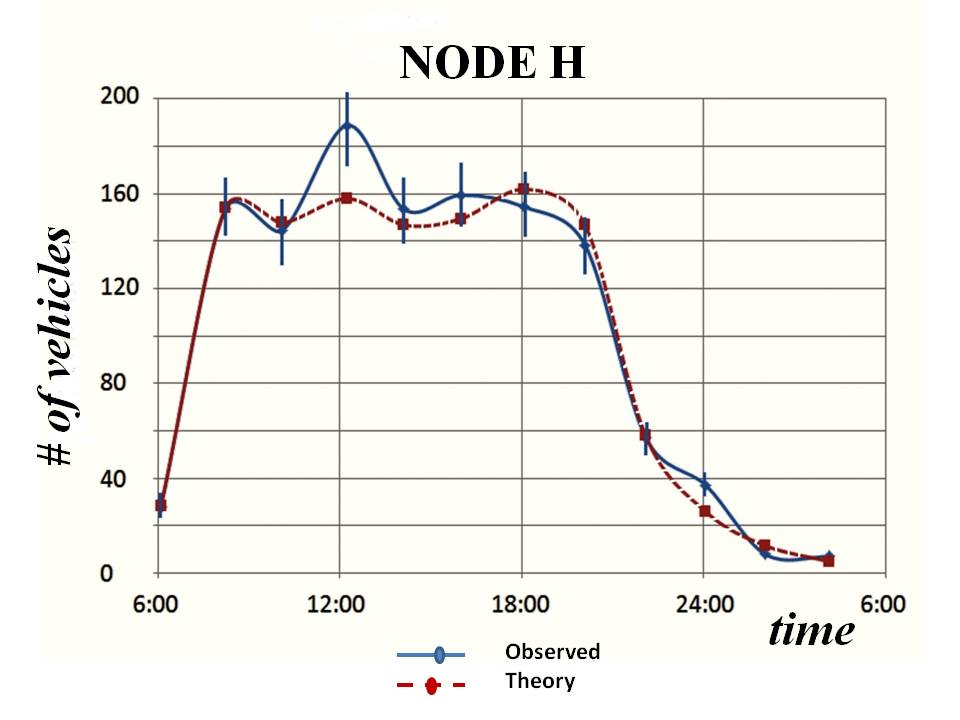}
\caption{\small{Variation of the flux (number of vehicles for every six 
minutes) going through the node $H$ along $22$ hours, 
beginning nearby 8:00 AM until 4:00 AM of the next day. Below the abscissa, 
the time in conventional notation. The error bars represent the dispersion of the 
empirical data collected in May 18, 2016. The hybrid model produces the dashed 
curve in red.}}
\label{p05}
\end{figure}

\begin{figure}[H]
\centering
\includegraphics[height=3.2in, width=3.4in]{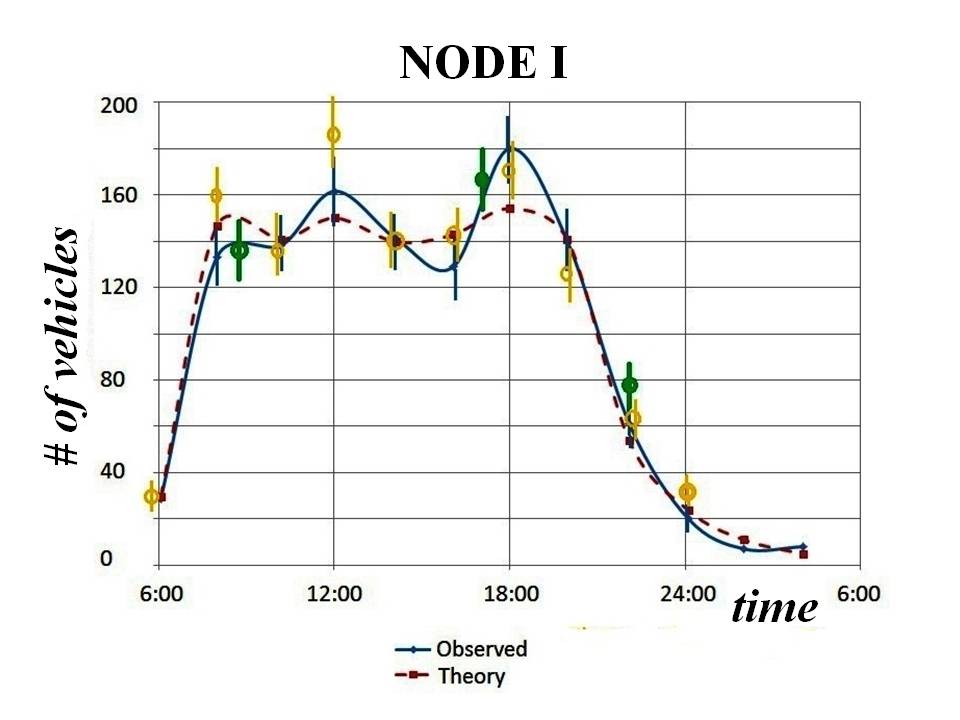}
\caption{\small{Variation of the flux (number of vehicles for every six 
minutes) going through the node $I$ along $22$ hours, 
beginning nearby 8:00 AM until 4:00 AM of the next day. Below the abscissa, 
the time in conventional notation. The error bars, in blue, represent the dispersion 
of the empirical data collected in May 18, 2016. The error bars, in green, represent the 
data collected in May 02, 2016. The error bars, in yellow, represent the data collected 
in May 19, 2016. The hybrid model produces the dashed curve in red.}}
\label{p06}
\end{figure}

\begin{figure}[H]
\centering
\includegraphics[height=3.2in, width=3.4in]{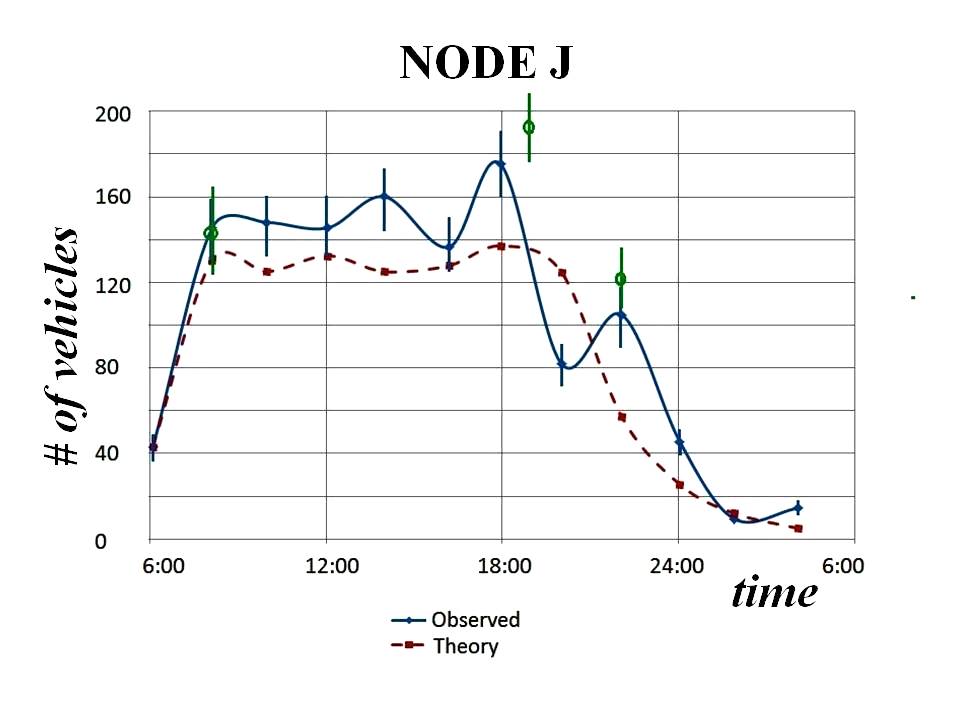}
\caption{\small{Variation of the flux of vehicles (number for every six 
minutes) going through the node $J$ along $22$ hours, 
beginning nearby 8:00 AM until 4:00 AM of the next day. Below the abscissa, 
the time in conventional notation. The error bars, in blue, represent the dispersion 
of the empirical data collected in May 18, 2016. The error bars, in green, represent 
the data collected in May 02, 2016. The hybrid model produces the dashed 
curve in red.}}
\label{p07}
\end{figure}

\begin{figure}[H]
\centering
\includegraphics[height=3.2in, width=3.4in]{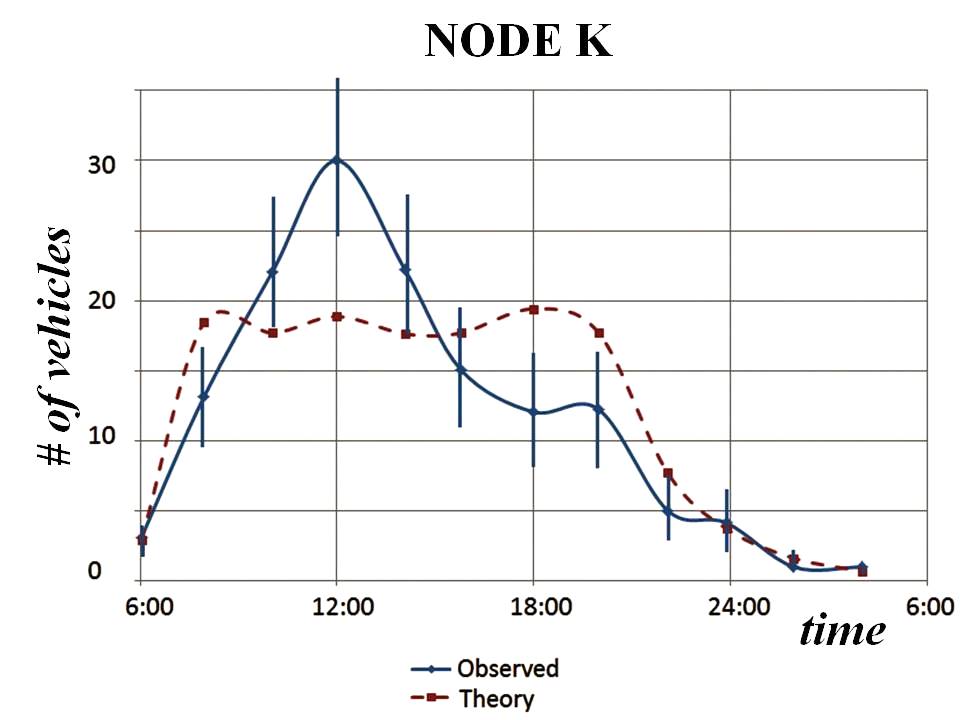}
\caption{\small{Variation of the flux (number of vehicles for every six 
minutes) going through the node $K$ along $22$ hours, 
beginning nearby 8:00 AM until 4:00 AM of the next day. Below the abscissa, 
the time in conventional notation. The error bars represent the dispersion of 
the empirical data collected in May 18, 2016. The hybrid model produces the dashed 
curve in red.}}
\label{p08}
\end{figure}

The rate of convergence to the quasi-stationary r\'{e}gime of the distribution 
of vehicles depends on $n$, and in accordance with each intersection it varies 
between 300 and 400. At the quasi-stationary r\'{e}gime the counted vehicles 
at the nodes fluctuate exceeding the Poisson dispersion. The dispersions are 
represented by the vertical bars in Figs. \ref{p01} to \ref{p08}.

We classified the daily data in four periods: 
(1) From 6:00 AM to 10:00 AM, when the influx starts with an almost zero flow 
of vehicles, then increasing along the first hours of the day. 
(2) From 10:00 AM to 08:00 PM the traffic roughly stabilizes, showing 
oscillations around some average value. The quasi-stationary r\'{e}gime 
is established and the difference between influx and outflux of vehicles 
is quite low. 
(3) From 08:00 PM to 04:00 AM of the next day, the vehicles begin leaving 
the network and an intense outflux, with almost no influx,  
is observed. 
(4) From 04:00 AM to 06:00 AM there is practically no circulation of 
vehicles within the network.

At the intersections with cameras, the vehicles standing idle at the 
semaphores or parked along the nearby sidewalks contributed to adjust the 
diagonal entries of matrices (\ref{s8}) and (\ref{s10}), while the vehicles 
crossing one intersection or changing direction (left or right) contributed 
to adjust the non-diagonal entries. At intersections without cameras the entries 
of the matrices were estimated and the numerical values were set such to get 
a best fit for the available data. The influx of vehicles within the network 
goes from node $P$ into the nodes $A,C,E,F,J,N$. 


In average, the number of vehicles at one intersection plus those in transit 
to the other intersections, at time $n$, is given by a component of the 
vector (\ref{s1}), with matrix (\ref{s8}) on the right hand side: 
$\mathbf{v}_{1}^{\dagger }\left( n\right) =\mathbf{v}_{0}^{\dagger }
\mathbb{M}_{eff}^{\left( 1\right) }\left( n\right) $. Notwithstanding, 
as the semaphores do not performed in good synchrony we opted to count the 
vehicles at each node independently of the others, so we projected out each 
component that was compared with the data. For instance, for node $k$ the 
average number of vehicles is
\begin{equation}
v_{1,k}\left( n_{k}\right) =\left( \mathbf{v}_{0}^{\dagger } \mathbb{%
M}_{eff}^{\left( 1\right) }\left( n_{k}\right) \right) 
\left( 
\begin{array}{c}
0_{1} \\ 
\vdots  \\ 
1_{k} \\ 
\vdots  \\ 
0_{15}%
\end{array}%
\right) .  \label{s5}
\end{equation}
At the beginning of the third period -- 8:00 PM to 04:00 AM --  
the traffic slows down because the vehicles begin to leave the sector,  
displaying an outflux from nodes $A,D,E,F,N,J$ and an influx 
towards the node $P$. The evolution of each node for an emptying 
network goes as 

\begin{equation}
v_{2,k}\left( n_{k}^{\prime },\bar{n}_{k}\right) =\left( \mathbf{v}%
_{0}^{\dagger }\mathbb{M}_{eff}^{\left( 1\right) }\left( \bar{n}%
_{k}\right) \right) \left(\mathbb{M}^{\left( 2\right) }\right)^{ n_{k}^{\prime
}} \left( 
\begin{array}{c}
0_{1} \\ 
\vdots  \\ 
1_{k} \\ 
\vdots  \\ 
0_{15}%
\end{array}%
\right) ,  \label{s5a}
\end{equation}
where $\bar{n}_{k}$ is now a fixed parameter, being the last number used for 
the previous period, whose value varies in the interval $\left( 2200,\  
2300\right) $, depending on each node $k$. The time is represented by 
$n^{\prime }_{k}$, and $\mathbb{M}^{\left( 2\right) }$ is the generator 
of the evolution for the third period. Depending on the node it refers, the 
\aspe last\aspd\ number $\bar{n}_{k}^{\prime }$ varies in the interval 
$\left( 1200,\ 1300\right) $. 

From 6:00 AM to 6:06 AM the Poisson dispersion is quite high because 
vehicles are not observed at some intersections while at others 
no more than six vehicles are counted (a dispersion around $45\%$). 
Comparatively, in the quasi-stationary r\'{e}gime the vehicles that 
cross the intersections of Caz\'{o}n avenue present an $8\%$
dispersion. 
\begin{equation}
\mathbb{M}^{\left( 1\right) }=\left( 
\begin{array}{ccccccccccccccc}
{ 0.002} & { 0.998} & { 0} & { 0} & { 0} & { 0}
& { 0} & { 0} & { 0} & { 0} & { 0} & { 0} & 
{ 0} & { 0} & { 0} \\ 
{ 0.531} & { 0.001} & { 0.468} & { 0} & { 0} & 
{ 0} & { 0} & { 0} & { 0} & { 0} & { 0} & 
{ 0} & { 0} & { 0} & { 0} \\ 
{ 0} & { 0.06} & { 0.045} & { 0.34} & { 0} & { %
0} & { 0} & { 0.555} & { 0} & { 0} & { 0} & { 0%
} & { 0} & { 0} & { 0} \\ 
{ 0} & { 0} & { 0.410} & { 0.010} & { 0.580} & 
{ 0} & { 0} & { 0} & { 0} & { 0} & { 0} & 
{ 0} & { 0} & { 0} & { 0} \\ 
{ 0} & { 0} & { 0} & { 0.145} & { 0.01} & { %
0.845} & { 0} & { 0} & { 0} & { 0} & { 0} & { 0%
} & { 0} & { 0} & { 0} \\ 
{ 0} & { 0} & { 0} & { 0} & { 0} & { 0.325} & 
{ 0.405} & { 0} & { 0} & { 0} & { 0} & { 0} & 
{ 0} & { 0.240} & { 0.030} \\ 
{ 0} & { 0} & { 0} & { 0.020} & { 0} & { 0.515}
& { 0.001} & { 0.464} & { 0} & { 0} & { 0} & { %
0} & { 0} & { 0} & { 0} \\ 
{ 0} & { 0} & { 0} & { 0} & { 0} & { 0} & 
{ 0.410} & { 0.180} & { 0.390} & { 0} & { 0} & 
{ 0.020} & { 0} & { 0} & { 0} \\ 
{ 0} & { 0.022} & { 0} & { 0} & { 0} & { 0} & 
{ 0} & { 0.143} & { 0.001} & { 0.834} & { 0} & 
{ 0} & { 0} & { 0} & { 0} \\ 
{ 0.321} & { 0} & { 0} & { 0} & { 0} & { 0} & 
{ 0} & { 0} & { 0.538} & { 0.017} & { 0.124} & 
{ 0} & { 0} & { 0} & { 0} \\ 
{ 0} & { 0} & { 0} & { 0} & { 0} & { 0} & 
{ 0} & { 0} & { 0.875} & { 0} & { 0.124} & { %
0.001} & { 0} & { 0} & { 0} \\ 
{ 0} & { 0} & { 0} & { 0} & { 0} & { 0} & 
{ 0} & { 0} & { 0} & { 0} & { 0} & { 0.010} & 
{ 0.990} & { 0} & { 0} \\ 
{ 0} & { 0} & { 0} & { 0} & { 0} & { 0} & 
{ 0.800} & { 0} & { 0} & { 0} & { 0} & { 0} & 
{ 0.006} & { 0.194} & { 0} \\ 
{ 0} & { 0} & { 0} & { 0} & { 0} & { 0} & 
{ 0} & { 0} & { 0} & { 0} & { 0} & { 0} & 
{ 0.900} & { 0.010} & { 0.090} \\ 
{ 0} & { 0} & { 0} & { 0} & { 0} & { 0.002} & 
{ 0} & { 0} & { 0} & { 0.007} & { 0} & { 0} & 
{ 0} & { 0} & { 0.991}%
\end{array}%
\right)  \ .
\label{s8}
\end{equation}

To describe the traffic from 8:00 PM to 4:00 AM, when the outflux of vehicles 
begins, Eq. (\ref{s5a}), we used the matrix (\ref{s10}), whose entries, also 
fixed by best. 
\begin{equation}
\mathbb{M}^{\left( 2\right) }=\left( 
\begin{array}{ccccccccccccccc}
{ 0.003} & { 0.997} & { 0} & { 0} & { 0} & { 0}
& { 0} & { 0} & { 0} & { 0} & { 0} & { 0} & 
{ 0} & { 0} & { 0} \\ 
{ 0.530} & { 0.001} & { 0.469} & { 0} & { 0} & 
{ 0} & { 0} & { 0} & { 0} & { 0} & { 0} & 
{ 0} & { 0} & { 0} & { 0} \\ 
{ 0} & { 0.030} & { 0.085} & { 0.320} & { 0} & 
{ 0} & { 0} & { 0.565} & { 0} & { 0} & { 0} & 
{ 0} & { 0} & { 0} & { 0} \\ 
{ 0} & { 0} & { 0.381} & { 0.040} & { 0.579} & 
{ 0} & { 0} & { 0} & { 0} & { 0} & { 0} & 
{ 0} & { 0} & { 0} & { 0} \\ 
{ 0} & { 0} & { 0} & { 0.140} & { 0.020} & { %
0.790} & { 0} & { 0} & { 0} & { 0} & { 0} & { 0%
} & { 0} & { 0} & { 0.050} \\ 
{ 0} & { 0} & { 0} & { 0} & { 0} & { 0.344} & 
{ 0.392} & { 0} & { 0} & { 0} & { 0} & { 0} & 
{ 0} & { 0.260} & { 0.004} \\ 
{ 0} & { 0} & { 0} & { 0.020} & { 0} & { 0.510}
& { 0.001} & { 0.469} & { 0} & { 0} & { 0} & { %
0} & { 0} & { 0} & { 0} \\ 
{ 0} & { 0} & { 0} & { 0} & { 0} & { 0} & 
{ 0.342} & { 0.188} & { 0.370} & { 0} & { 0} & 
{ 0.100} & { 0} & { 0} & { 0} \\ 
{ 0} & { 0.010} & { 0} & { 0} & { 0} & { 0} & 
{ 0} & { 0.045} & { 0.001} & { 0.944} & { 0} & 
{ 0} & { 0} & { 0} & { 0} \\ 
{ 0.320} & { 0} & { 0} & { 0} & { 0} & { 0} & 
{ 0} & { 0} & { 0.469} & { 0.090} & { 0.120} & 
{ 0} & { 0} & { 0} & { 0.001} \\ 
{ 0} & { 0} & { 0} & { 0} & { 0} & { 0} & 
{ 0} & { 0} & { 0.875} & { 0} & { 0.124} & { %
0.001} & { 0} & { 0} & { 0} \\ 
{ 0} & { 0} & { 0} & { 0} & { 0} & { 0} & 
{ 0} & { 0} & { 0} & { 0} & { 0} & { 0.010} & 
{ 0.990} & { 0} & { 0} \\ 
{ 0} & { 0} & { 0} & { 0} & { 0} & { 0} & 
{ 0.800} & { 0} & { 0} & { 0} & { 0} & { 0} & 
{ 0.006} & { 0.194} & { 0} \\ 
{ 0} & { 0} & { 0} & { 0} & { 0} & { 0} & 
{ 0} & { 0} & { 0} & { 0} & { 0} & { 0} & 
{ 0.960} & { 0.010} & { 0.030} \\ 
{ 0} & { 0} & { 0} & { 0} & { 0} & { 0} & 
{ 0} & { 0} & { 0} & { 0} & { 0} & { 0} & 
{ 0} & { 0} & { 1}%
\end{array}%
\right)  \ .
\label{s10}
\end{equation}

The numerical entries of matrices $\mathbb{M}^{\left( 1\right) }$ 
and $\mathbb{M}^{\left( 2\right) }$ were obtained by minimizing  
$\chi ^{2}$, for influx and outflux, to get the best fit \cite{bevinrobin}, 
\be
\chi _{influx}^{2}=\sum_{j=1}^{8}\sum_{k=1}^{8}\frac{\left(
v_{1,k,j}^{obs}-v_{1,k,j}^{theor}\right) ^{2}}{v_{1,k,j}^{obs}}\ ,
\label{chiin}
\ee
and
\be
\chi _{outflux}^{2}=\sum_{j=1}^{8}\sum_{k=1}^{4}\frac{\left(
v_{2,k,j}^{obs}-v_{2,k,j}^{theor}\right) ^{2}}{v_{2,k,j}^{obs}}\ ,
\label{chiout}
\ee
where the subscript $j$ stands for the eight intersections with cameras and $k$ for 
the six minutes sample at every two hours. For the influx counting we had 62 effective 
observations out from the 64 possible ones, while for the outflux all 32 observation 
were effective. For the influx transient and the quasi-stationary r\'{e}gime (06:00 AM 
to 08:00 PM) we have considered: (1) the normalization of each row of the 
matrix and, (2) the effects of the environment on the network, namely, 
in Eq. (\ref{s2}) we set $\mathcal{N}=1970$ and the most adequate function is
\begin{equation}
f\left( n\right) =\left( 20.0+0.0576\ n\right) \cos \left( \frac{\pi n}{%
275.0+0.055\ n}\right) ,  \label{s6}
\end{equation}
which depicts the oscillations with varying amplitudes and periods. At the nodes 
where the traffic was monitored the experimental fluxes and those resulting from the 
hybrid model are shown in the Figs. \ref{p01} to \ref{p08}. The vertical axis 
stands for the number of vehicle counted every six minutes and in the horizontal 
axis the time considering $ 22.5\ s$ as a unit. The experimental data are from 
three different days: May 18, 2016 (blue solid line and error bars in the graphs) 
present in all the eight nodes, Figs. \ref{p01} to \ref{p08}. The data from 
April 30 and May 02, 2016 (green error bars in the graphs) are for the nodes 
$F,G,I$ and $J$, shown in Figs. \ref{p03}, \ref{p04}, \ref{p06} and \ref{p07}. 
The data from May 19, 2016 (yellow error bars in the graphs) are for the node 
$I$ only, shown in Fig. \ref{p06}. The blue solid curve that links the experimental 
points were obtained by interpolation using the spreadsheet\emph{ Microsoft Excel}. 
The curves resulting from our model are represented by red dashed lines. In the 
quasi-stationary r\'{e}gime the hybrid model presents oscillations with, in general, 
three resolved peaks, although the empirical data show two or three peaks. 
Nevertheless the curves of the hybrid model stand for the best fit for the traffic 
flux represented by the distribution of vehicles at all the nodes. Our assessment 
is that the model and the theoretical approach we embraced describe satisfactorily 
the traffic complexity within our chosen urban sector.
%
\section{Additional analysis}
\begin{figure}[htbp]
\centering
\includegraphics[height=4.0in, width=5.0 in]{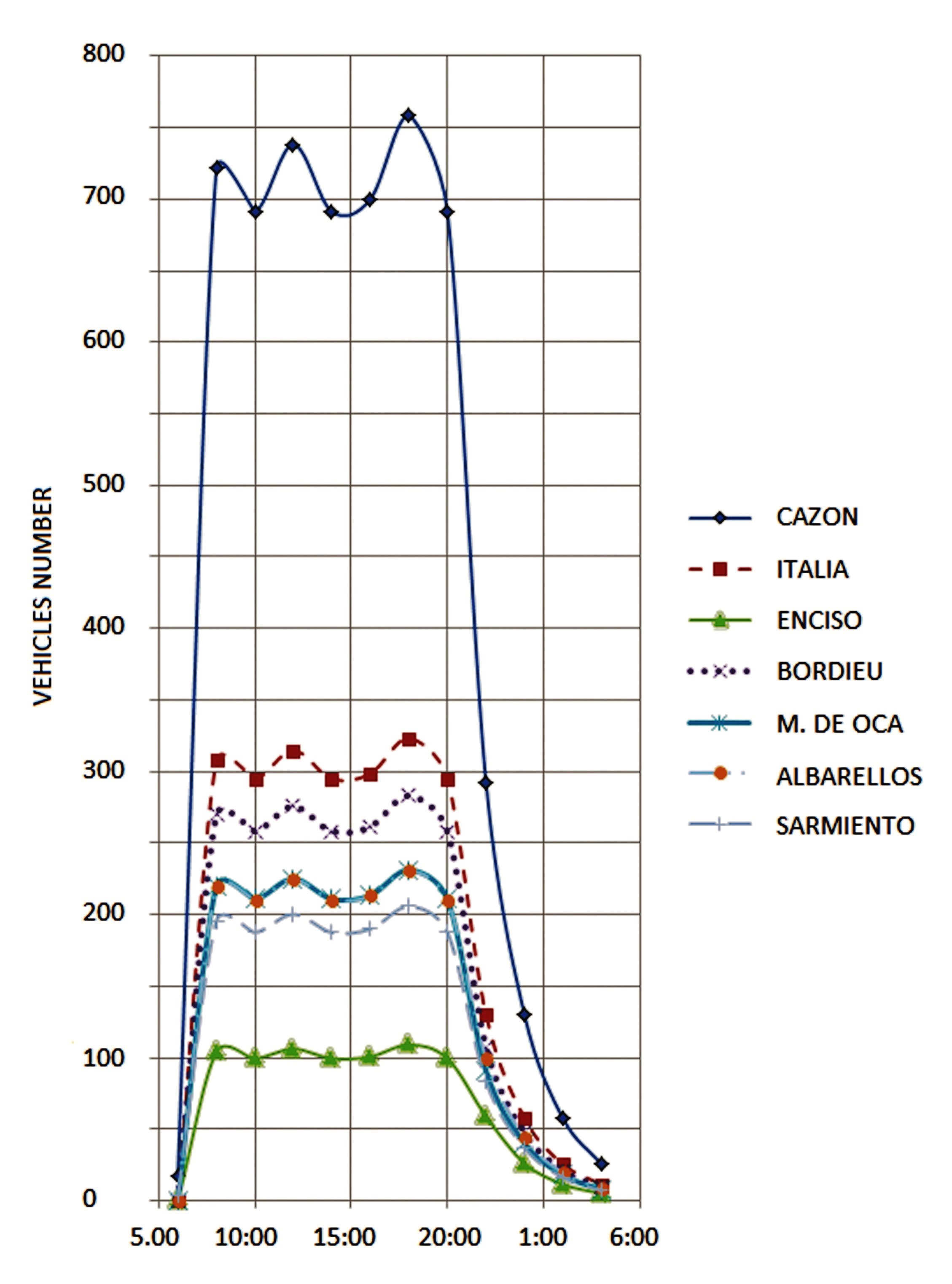}
\caption{\small{The vertical axis represents the flux (number of vehicles 
during six minutes), according to the model, for several one-way arteries, or 
two-way as Caz\'{o}n, Italia and Enciso. 
.}}
\label{p12}
\end{figure}

%
As one of our objectives is to know the number of vehicles circulating 
through the arteries of the city sector at different moments of the day, we adopted 
$\mathbb{M}^{\left( 1\right)}$ and $\mathbb{E} \left( 1 \right)$ as the 
time-zero matrices, i.e., when the vehicular influx begins. In Fig. \ref{p12} 
we display the evolution of the traffic flux calculated as the number of vehicles 
per six minutes along the considered arteries. 
The dots in the figure stand for $\mathbf{v}_{1}\left( n_0 j\right)$ 
($n_0 = 320$) with $j=1,2,...,6$ for times 8:00 AM, 10:00 AM, 12:00 AM, 02:00 PM, 
04:00 PM, 06:00 PM and $\mathbf{v}_{2}\left( n_{0}^{\prime } j\right) $ 
($n_{0}^{\prime }=320$) and $j=1,2,...,5$ for times 08:00 PM, 10:00 PM, 12:00 PM, 
02:00 AM, 04:00 AM. 

We hold that modeling serves to assist the traffic controllers by proposing 
procedures for optimizing the vehicular circulation by diminishing the 
probability of jams, car accidents and reducing the average travel time 
between two locations. In particular, a model study can also recommend how to 
better set the timing of the semaphores lights in the intersections, even in 
those where cameras are not installed, and to synchronize conveniently their 
delay times along the main arteries. 

Caz\'{o}n, a two-way avenue, possesses semaphores and cameras at the five 
intersections comprised in the chosen sector. The green light duration at 
each intersection is $72\ s$ while the red light stays on for $38\ s$. As 
the considered segment of Caz\'{o}n avenue has a length of $520\ m$, by 
starting at intersection $J$ until crossing $F$ a vehicle could run through 
it, ideally, in $\approx 47\ s$ at an average speed of 40 $km/h$ ($\approx 
11.1\ m/s$), without stopping at a red light. Nevertheless, what was observed 
is that within the period from 10:00 AM to 08:00 PM (the quasi-stationary 
r\'{e}gime), it takes around $100\ s$ for a vehicle to go along that section 
with a unique stop at a red light. So the effective time 
for a vehicle to run along Caz\'{o}n goes from the ideal $47\ s$ to the 
real $\approx 100\ s$, which is due to the jam caused by buses stopping at bus 
stops. So the average speed of a running vehicle is reduced to $30\ km/h$, and 
considering the idle time at one red light, this effectively goes down to 
$\approx 19\ km / h $ ($5.2\ m / s$). 

Even so, an improvement in the traffic circulation can be made through a simple 
analysis. At the quasi-stationary r\'{e}gime we considered the ratios for the flux 
of vehicles for three major adjacent intersecting streets of Caz\'{o}n: $I$ 
(for Caz\'{o}n-Bordieu), $H$ (for Caz\'{o}n - Montes de Oca) and $G$ 
(for Caz\'{o}n - Albarellos),  
\begin{equation}
R_{\alpha }=\frac{\text{Flux of vehicles along Caz\'{o}n}}{\text{Flux of 
vehicles along }\alpha }\ ,
\label{ratiocazon}
\end{equation}
where $\alpha $ can be $I,H$ or $G$. In average, the calculated numerical 
results are $R_{I} \approx 2.7,\quad R_{H}\approx 3.2,\quad R_{G} 
\approx 3.7$, that can also be estimated from the curves in Fig. 
\ref{p12}. With the aim of improving the traffic circulation along 
Caz\'{o}n avenue and without much impairment to the traffic flux of the 
intersecting arteries, we have assumed the lowest ratio $R=2.7$ for all three 
intersections. The measured ratio of times for which the semaphore lights remain 
green is $T_{\text{Caz\'{o}n}}/T_{\text{intersection}}=72\ s/38\ s \approx 1.9$. 
Considering $R=2.7$, our model hints to an improvement of the circulation 
if the traffic controller keeps the $72\ s$ green light time for Caz\'{o}n avenue 
and reduces the green light time for the intersections to $72/2.7\approx 27\ s$. 
The reduction from $38\ s$ to $27\ s$ for one single stop at a red light in 
Caz\'{o}n provides around $30\%$ economy of fuel consumption of an idle vehicle, 
permitting thus a significant reduction of CO$_{2}$ and pollutants emissions, 
besides benefiting drivers and passengers with a gain in time.
%
\section{Summary and conclusions}
%
In Part I of our research \cite{ogm1}, we have proposed three models 
that describe ideally the traffic flow. In the present Part II we  
used the proposed models to analyze and emulate a real situation for the 
urban traffic. We counted the number of vehicles from tapes disposed to us 
by the traffic controllers, regarding the circulation of vehicles in that 
particular sector of the city, shown in Fig. \ref{sector}. The recording cameras 
were disposed at eight intersections. The data were obtained from a sample time 
of six consecutive minutes every two hours during a full 24 hours day and they 
were used to: (1) calculate the distribution of the vehicles moving along the arteries 
with the respective bars of errors, and (2) to find out an adequate model, 
that turned to be essentially a hybrid one, which combines the open network 
model with the nonlinear one, both presented in Part I. This hybrid model 
is dynamical, i.e., it describes the time evolution of the traffic within 
the network continuously along a whole day. We have also compared the 
model -- constructed from data of one single day -- with the data obtained 
from different days. To our judgment this model is sound within the 
experimental errors, since it is able to describe satisfactorily the 
dynamical traffic trend. In order to optimize the traffic flow the model 
hinted for a change in the semaphores red light time duration. Finally, we estimate 
that the model is scalable, and as such it can be extended to larger networks.
%
\begin{acknowledgments}
SSM thanks the CNPq, a Federal Brazilian Agency, for financial support.
\end{acknowledgments}
%

%
\end{document}